\documentclass{vgtc}                          

\ifpdf
  \pdfoutput=1\relax                   
  \usepackage{graphicx}                
  \DeclareGraphicsExtensions{.pdf,.png,.jpg,.jpeg} 
\else
  \usepackage{graphicx}                
  \DeclareGraphicsExtensions{.eps}     
\fi%

\graphicspath{{figures/}{pictures/}{images/}{./}} 

\usepackage{microtype}                 
\PassOptionsToPackage{warn}{textcomp}  
\usepackage{textcomp}                  
\usepackage{mathptmx}                  
\usepackage{times}                     
\usepackage{cite}                      
\usepackage{tabu}                      
\usepackage{booktabs}                  
\usepackage{verbatim}

\onlineid{0}

\vgtccategory{Research}

\vgtcinsertpkg

\title{RuleVis: Constructing Patterns and Rules for Rule-Based Models}

\author{David Abramov$^1$, Jasmine Otto$^1$, Mahika Dubey$^1$, Cassia Artanegara$^1$, \\Pierre Boutillier$^2$, Walter Fontana$^2$, Angus G. Forbes$^1$ \\ \\
\parbox{2.8in}{\scriptsize \centering $^1$Department of Computational Media \\ University of California, Santa Cruz \\ \{dabramov,jtotto,mahika,cartaneg,angus\}@ucsc.edu}
\parbox{2.8in}{\scriptsize \centering $^2$Department of Systems Biology\\ Harvard Medical School \\ \{pierre\_boutillier,walter\_fontana\}@hms.harvard.edu}
}

\teaser{
  \centering
  \fbox{\includegraphics[width=6in]{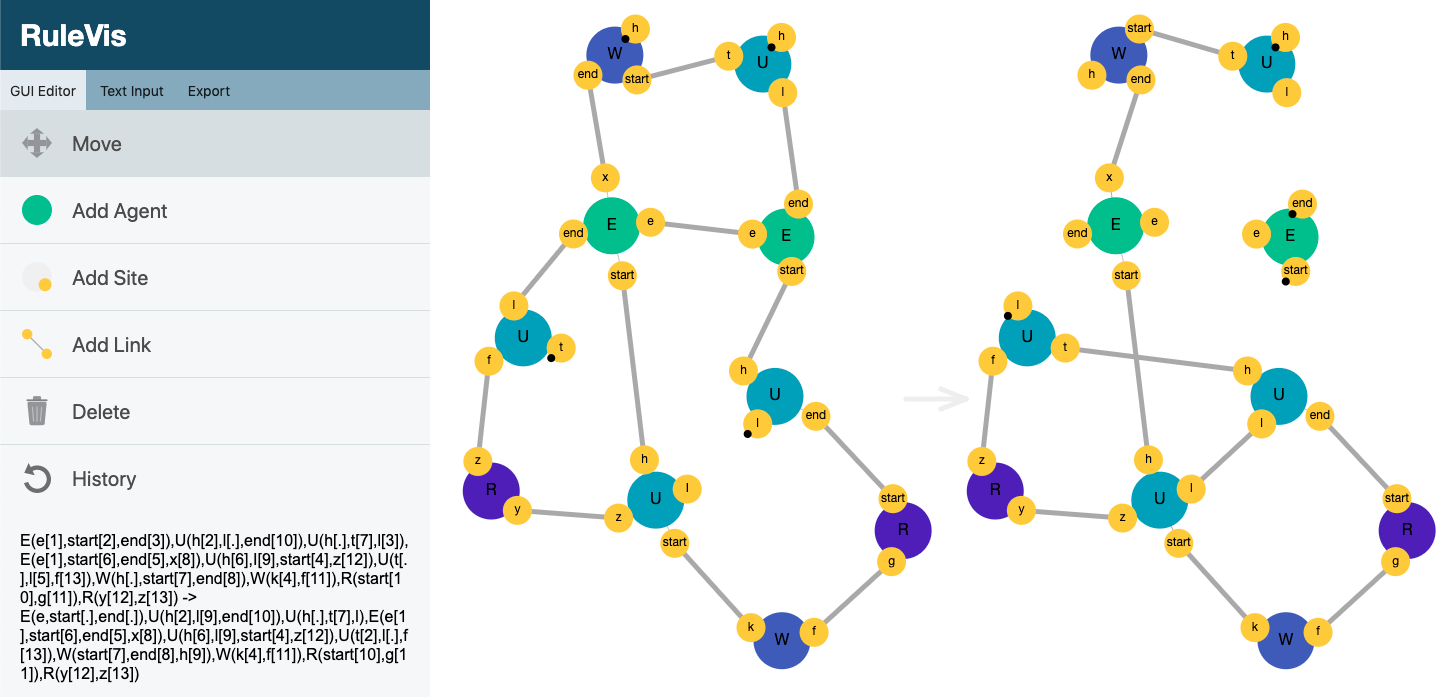}}
  \caption{A screenshot of the \textit{RuleVis} interface, depicting a user in the process of building a rule, consisting of two patterns that describe the `before' and `after' states of the system (separated by an arrow). The editor panel (left) enables users to add and delete agents, sites, and links in the visualization in the display panel (right). The text and visualization are mirrored, and changes made to one representation are immediately reflected in the other.}
  \label{fig:teaser}  
}

\abstract{We introduce \textit{RuleVis}, a web-based application for defining and editing ``correct-by-construction'' executable rules that model biochemical functionality, which can be used to simulate the behavior of protein-protein interaction networks and other complex systems. Rule-based models involve emergent effects based on the interactions between rules, which can vary considerably with regard to the scale of a model, requiring the user to inspect and edit individual rules. \textit{RuleVis} bridges the graph rewriting and systems biology research communities by providing an external visual representation of salient patterns that experts can use to determine the appropriate level of detail for a particular modeling context. We describe the visualization and interaction features available in \textit{RuleVis} and provide a detailed example demonstrating how \textit{RuleVis} can be used to reason about intracellular interactions.

} 


\begin{document}


\firstsection{Introduction \& Background}
\maketitle

A central challenge in characterizing complex systems, such as cells, is to understand how a multitude of heterogeneous agents interact to generate coherent behavior~\cite{kitano2002computational}. The individual components and their continuous interactions often result in an overwhelming complexity, frustrating any initial insight needed for guiding the construction of a model~\cite{deeds12}. Rule-based models simplify the process of constructing meaningful models and make it easier to update them if underlying assumptions change, and to evaluate the consequences of these updates~\cite{chylek2015modeling,wilson2015kappa}. In this paper, we present a new component of the Kappa ecosystem called \textit{RuleVis}. Kappa is a rule-based modeling language that uses a context-free grammar to formalize biochemical notation into a standard format for use in computational modeling. It enables the construction, simulation, and analysis of models that can capture the combinatorial complexity observed in molecular processes, allowing researchers to reason about the statistical dynamics
induced by the interaction of heterogeneous agents~\cite{kappamanual}.

\textit{RuleVis} enables researchers to a) \textit{construct} Kappa rules from scratch, b) \textit{import} existing rules, c) \textit{edit} rules, and d) \textit{export} the rules into other Kappa applications for simulation and analysis~\cite{boutillier2018kappa}, as well export high-quality figures that can be used in publications or for educational purposes. The GUI editor allows for construction and editing of rules without explicit knowledge of the underlying textual Kappa syntax. The textual editor also enables construction and editing tasks and is essential for importing rules initially defined using other software and for exporting rules to interface with other Kappa applications. Fig.~\ref{fig:teaser} shows the \textit{RuleVis} interface, demonstrating in-progress rule creation by a user.


There is a rich history of interactive visualization tools to support system modeling across many domains~\cite{biovisDang2,Dang2016a_IEEEEuroVis,goel2009structure,kitano2005using,kolpakov2006biouml,le2009systems,moody2008evaluating,MurrayBMC2017}. Popular platforms for accessing biological pathways, such as Reactome~\cite{fabregat2015reactome}, provide a curated database of interactive pathway diagrams that summarize the behavior of known biological processes. Cytoscape~\cite{shannon2003cytoscape,kohl2011cytoscape} is a network visualization platform that generates effective layouts for large biological networks. Yet, as Gehlenborg et al.~\cite{gehlenborg2010visualization} articulate, it remains an ongoing challenge in biological data visualization to create effective visualizations that provide insight into intrinsically complex data.

\begin{figure}
    \centering
      \includegraphics[width=0.95 \columnwidth]{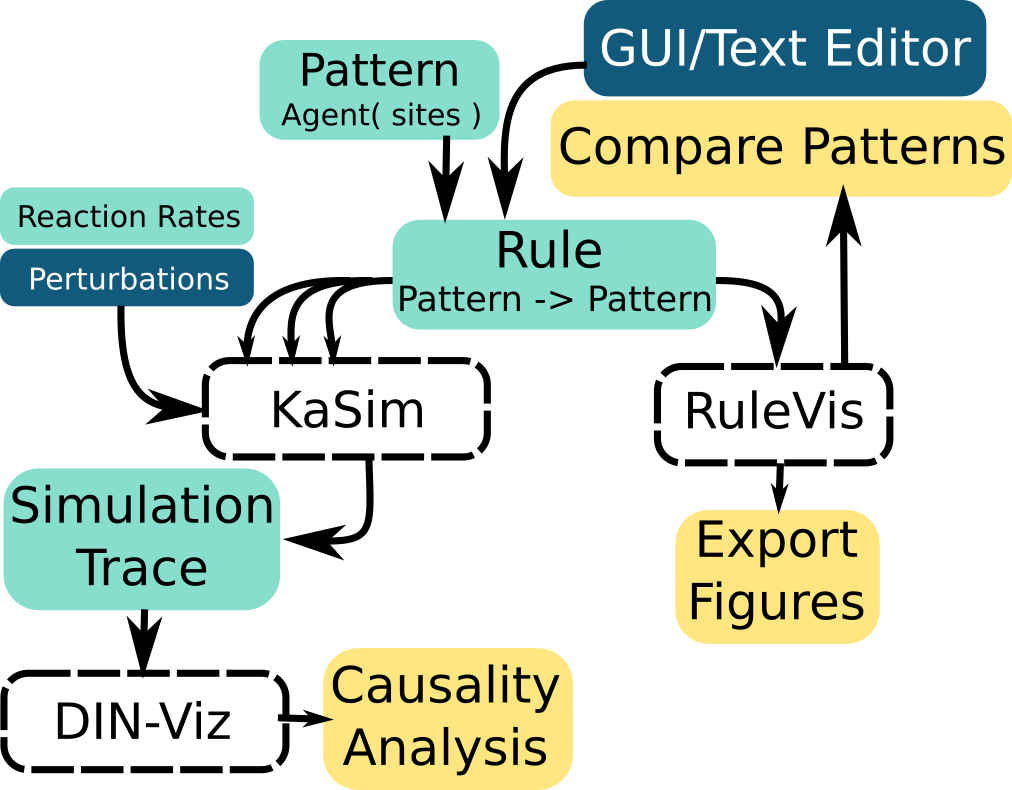}
      \caption{\textit{RuleVis} is situated within the Kappa Language ecosystem, providing graphical editing capabilities for constructing individual patterns and rules used in different Kappa applications. In this figure, individual applications are indicated by the boxes with dashed borders. These applications support a range of analysis tasks (gold), data processing tasks (navy blue), and visualization outputs (teal).}
    \label{ecosystem}
\end{figure}

Systems biology is a source of complex data, which requires specialized semantics such as rule-based modeling. Smith et al.'s \textit{RuleBender}~\cite{smith2012rulebender} provides a visual interface for simulating, visualizing, and editing rules in the BioNetGen Language~\cite{harris2016bionetgen}. More recently, Sekar et al.~\cite{sekar2017automated} introduce an alternative representation of rules that includes a ``tunable compression pipeline'' to generate more compact layouts. \textit{RuleVis} visualizes patterns and rules written in the Kappa language, and also facilitates the interactive construction of new rules via a visual interface, leveraging a user's visual intuition when generating Kappa expressions. \textit{RuleVis} was designed to interface with other software in the Kappa ecosystem, as shown in Fig.~\ref{ecosystem}, such as \textit{KaSim}, a stochastic simulator that records the system's state as it moves forwards in time, including the rates of activation of each rule~\cite{boutillier2018kappa,forbes2017dynamic}.

In addition to facilitating the construction of patterns and rules to support execution of rule-based models, the \textit{RuleVis} interface supports the visual externalization of a user's mental model of how the salient features of a given complex system are interconnected, supporting communication and collaboration. For example, while biological pathways represent our current understanding of a given intracellular process, biologists may need to reason about the relevance of a pathway for a particular context, or determine if recent discoveries impact a current research question. \textit{RuleVis} aims to bridge the gap between computational and experimental biologists~\cite{danos2012graphs} by making rules visible and editable. Researchers can explicitly describe their model of the pathway's salient elements, quickly produce high-quality figures, and then invite others to validate the model. Molecular biologists may not be familiar with rule-based modeling, and our interface hides some of the more technical details of graph-rewriting, enabling the user to focus on the correctness of the model, and to make sure that it accurately represents relevant aspects of the system under investigation. 

Using \textit{RuleVis} in the context of \textit{KaSim}, a researcher can iterate between possible solutions, and modify rules to see how that changes the output of execution. This iterative approach also lends itself to the classroom, and an ongoing project integrates rule-based models into introductory systems biology courses. \textit{RuleVis} simplifies the construction of rules, and makes it possible for students to quickly create and test their own models.

\section{The \textit{RuleVis} Application}

\subsection{\textit{Kappa} Patterns and Rules} 



\textit{RuleVis} is an interactive tool for visualizing and editing \textit{rules} made out of \textit{patterns}, which comprise executable models of complex systems. Rule-based modeling languages such as Kappa represent the structures organizing macromolecular agents as \textit{patterns}. In this formulation, individual agents are analogous to atoms. Just as atoms have physical properties that allow them to bind into molecules with other atoms, agents have biochemical properties that allow them to form larger patterns with other agents. Each agent has ``sites'' with some internal state, which can bind to other sites on other agents, forming a pattern. These links between sites are referred to as ``edges,'' which can only be placed once per site and cannot connect to a site on the same agent. Just as molecules undergo chemical reactions that cause their bonds to change, rearranging the atoms, a Kappa \textit{rule} defines how a pattern can change its internal state or be transformed in the presence of other patterns~\cite{boutillier2018kappa}. Given appropriate sets of rules, the Kappa graph rewriting engine is able to generate coherent oscillatory behaviors that emulate complex biological behaviors~\cite{forbes2017dynamic}.

\begin{figure}
    \centering
      \includegraphics[width=0.98\columnwidth]{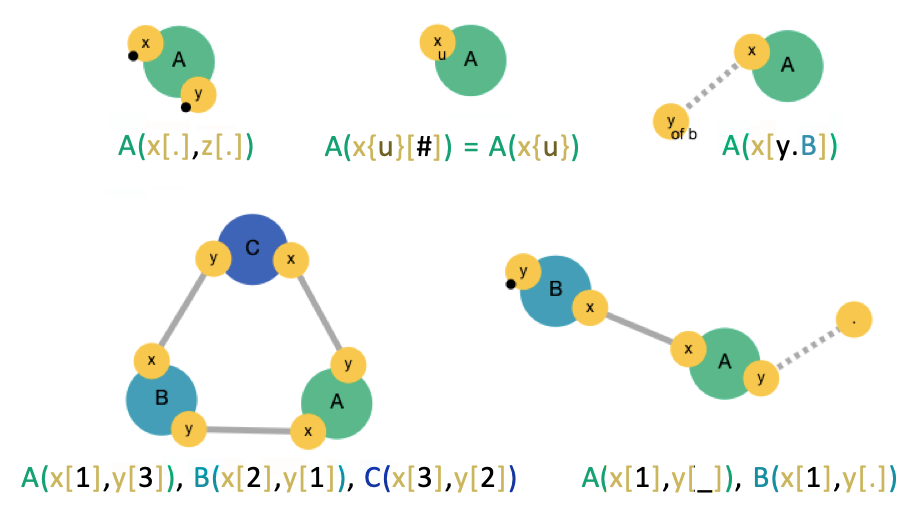}
      \caption{Examples of simple Kappa patterns represented both in text and their associated visual form. Here we show a range of sites in various unbound and linked configurations.} 
    \label{basic-patterns}
\end{figure}

More explicitly, a \textit{pattern} is a site graph where multiple types of the same agent are allowed to occur, but each site can have at most only one edge~\cite{danos09,danos10}. In the Kappa language, an expression consists of its type, or name, followed by a list of comma-separated site names, with the state of a site specified after its name. Sites can be in one of two states: an internal state (usually representing a modification state) enclosed in curly brackets, and a link state (usually representing a bond), which is specified as a non-negative integer written between square brackets. This value determines when an edge is present between two sites. The link state of a site that is unbound (free) is indicated with a period in square brackets. Fig.~\ref{basic-patterns} demonstrates this language structure with a few examples that map a textual representation of a pattern to its visual representation. 

\begin{figure}[t!]
    \centering
      \includegraphics[width=3.25in]{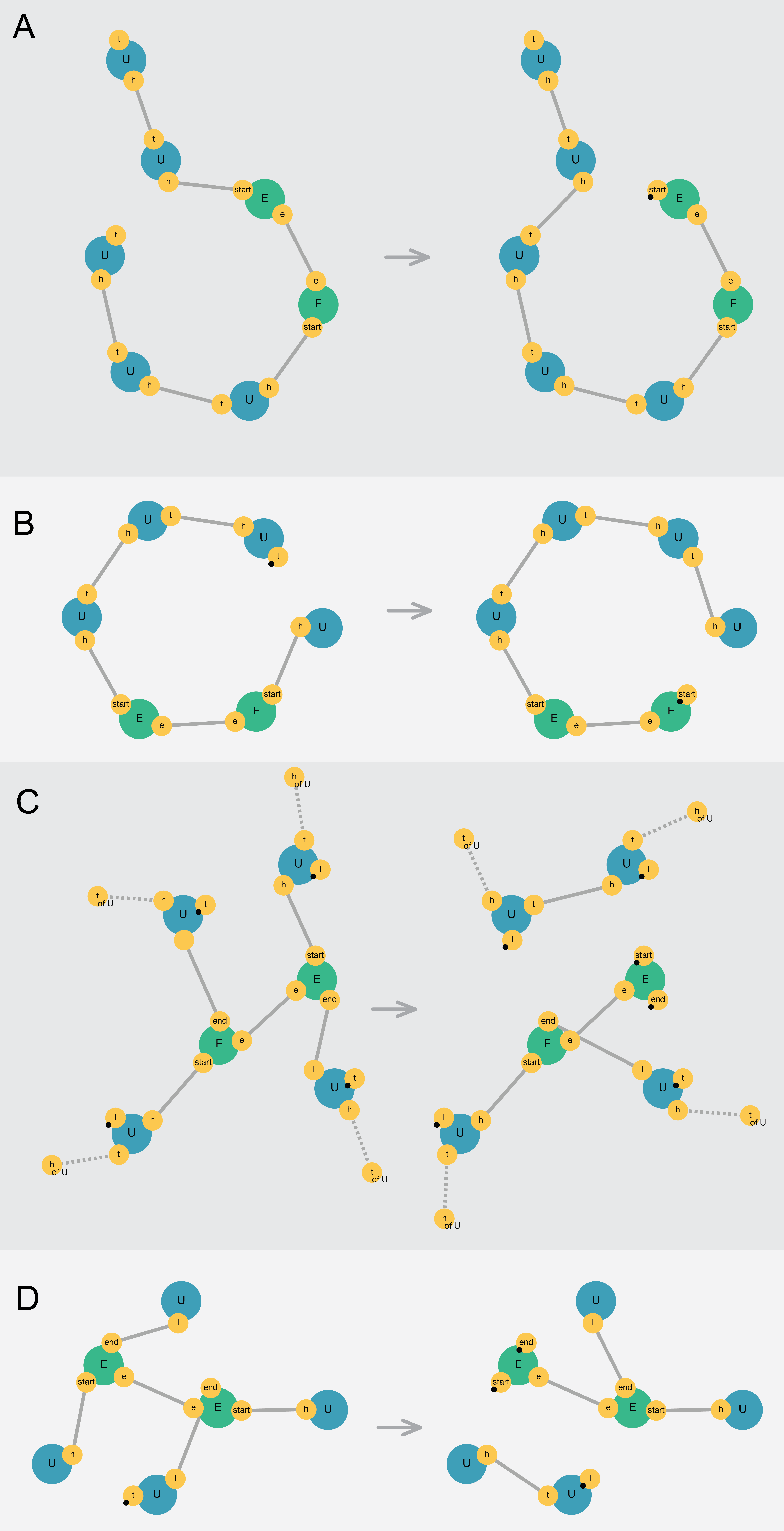}
      \caption{Example rules visualizing `transfer of chains,' where consecutively linked agents (chains) are moved to a different agent in the graph. (A) shows the transfer of a chain of two \textit{U} agents from one agent \textit{E} onto the end of another \textit{U} agent chain connected to a different agent \textit{E}. (B) shows the transfer of a chain of length 1 onto a chain of length 3. (C) shows the transfer of one chain onto another while preserving the information about the tail ends of their chains. (D) shows a rule for concatenating two chains of arbitrary length \cite{kappamanual}.} 
    \label{fig:example-patterns}
\end{figure}

An expression of a \textit{rule} in the Kappa language consists of two patterns separated by a right arrow, providing an abstraction of protein-protein interactions, where specificity of the molecular interactions is not important for the overall model~\cite{kappamanual}. Fig.~\ref{fig:example-patterns} depicts visual representations of rules that transform a pattern on the left into the pattern on the right. The number of agents on the left and the right side of the expression must match in order for the expression to parse correctly. Every rule contains a fixed number of agents on each side, each of which has the same number of sites on each side. However, bond state can vary freely between sides if no one site possesses more than one bond on either side. \textit{RuleVis} presents both the textual representation and the visual representation of each rule to the user, either of which they can manipulate to receive real-time feedback. Any valid Kappa expression can be typed into our application to visualize it instantly, and any interactive change to the visual graph will likewise update the textual representation. \textit{RuleVis} includes a set of constraints that enforce rules to be ``correct-by-construction,'' whereby only valid expressions can be visualized. For example, when using the GUI editor, invalid operations highlight links and/or sites in red. Likewise, expressions typed in text editor will not update the visualization until the syntax entered explicitly fulfills the requirements of Kappa's context-free-grammar.

\subsection{Application Details}

The \textit{RuleVis} application consists of an editor panel and a display panel. The editor panel contains three tabs. The first tab provides a graphical user interface for creating and editing graphs (as shown in Fig.~\ref{fig:teaser}). The second tab provides a text editor where rules can be created or edited using Kappa language syntax, and then immediately visualized in the display panel. The third tab provides options for saving data in different formats. \textit{RuleVis} supports downloading an SVG file of the rule diagram (which can be loaded into image editing software such as Inkscape or Adobe Illustrator), as well as exporting a JSON file that contains the necessary metadata to recreate a rule visualization for further editing in a later session. The display panel takes up the majority of the screen real estate, and produces a graphical representation of Kappa patterns and rules. Users can interactively edit these rules and update their layout on demand.

\begin{figure}
    \centering
      \includegraphics[width=.9\columnwidth]{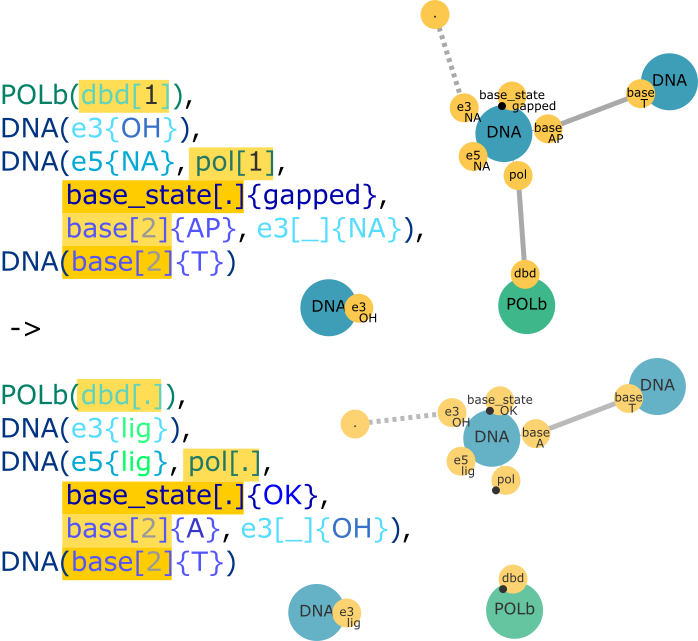}
      \caption{Example of a rule with many changing agents and multiple site states. There are unbound sites with no links, sites bound to the sites of other agents, and one virtual bond, whose identity is not necessarily preserved in the context of the rule. The colors in the text syntax and network visualization are aligned, where blue and green correspond to agents \textit{DNA} and \textit{POLb} and the yellow text corresponds to the sites.}
      
    \label{fig:complex-rule}
\end{figure}

Every Kappa rule, consisting of two pattern expressions--- the `before' and `after'--- is visualized using two copies of a merged expression graph, so that initially two copies of a node remain in the same position on both sides of the rule. Every expression is visualized as a graph with `agent' nodes directly connected to one or more `site' nodes, and where site nodes for one agent can bind to the sites nodes for another agent. Agents are displayed as larger circles, while each of their associated sites is visualized as smaller yellow circles. Pressing the `Add Agent' and `Add Site' buttons (in the editor panel) creates new node elements in the display panel, and produces a text box for entering a name for the new node. A default color palette distinguishing agents with different names, and \textit{RuleVis} automatically matches the colors of agents given the same names. A solid gray line is used to indicate that agents are bound together (via their sites). Anonymous bonds, in which sites are bound, but not directly associated with a specified agent, are visualized with a dotted gray line.  Sites can also be in an unbound state, which is indicated with a black dot placed on their edge. Fig.~\ref{fig:complex-rule} shows an example visualization that includes each of these visual encodings, demonstrating the mappings between the textual representation of the rules and the visual representation. 

Agents and sites can be interactively re-positioned in the display panel. Our layout constrains the position of any agent, site, or link to within the extents of the canvas, preventing nodes from falling out of view. Agents, sites, and links can be deleted while still maintaining the element hierarchy. To preserve the balancing of both sides of the rule, as well as speed up the creation of the visualization, placing an agent or site in the canvas populates both the right and left hand side of the expression, and changes can be made on a single side using a `delete' tool to carve out differences between the patterns on either side of the rule. We also chose to geometrically link together the left and right hand sides of the expression, so that, for example, when a node on one side is moved, the corresponding node on the other side of the expression also moves simultaneously. This makes it easy to map between the two sides of the expression, and reflects the balanced format in which chemical expressions are typically written.



\subsection{Technical Implementation}
\textit{RuleVis} is a lightweight web application written in Javascript that runs in any browser. We use D3.js to manage the interaction with the dynamic SVG canvas~\cite{bostock2011d3} and the the Cola.js library to implement the constraint-based layout of our visualization~\cite{Dwyer_COLA}. We use Lahodiuk's implementation\footnote{\url{https://github.com/lagodiuk/earley-parser-js}} of an Earley parser in Javascript to implement the  context-free grammar specified by Kappa documentation,  benefiting from the modularity of mathematical notation~\cite{earley1970efficient}. Besides providing live feedback to experienced rule-writers, the manipulation of pre-existing example rules is a valuable technique for learning systems modelling with Kappa. For other visual expressions, this would be impossible without a robust serialized representation.




\begin{figure}[t!]
    \centering
      \includegraphics[width= 1\columnwidth]{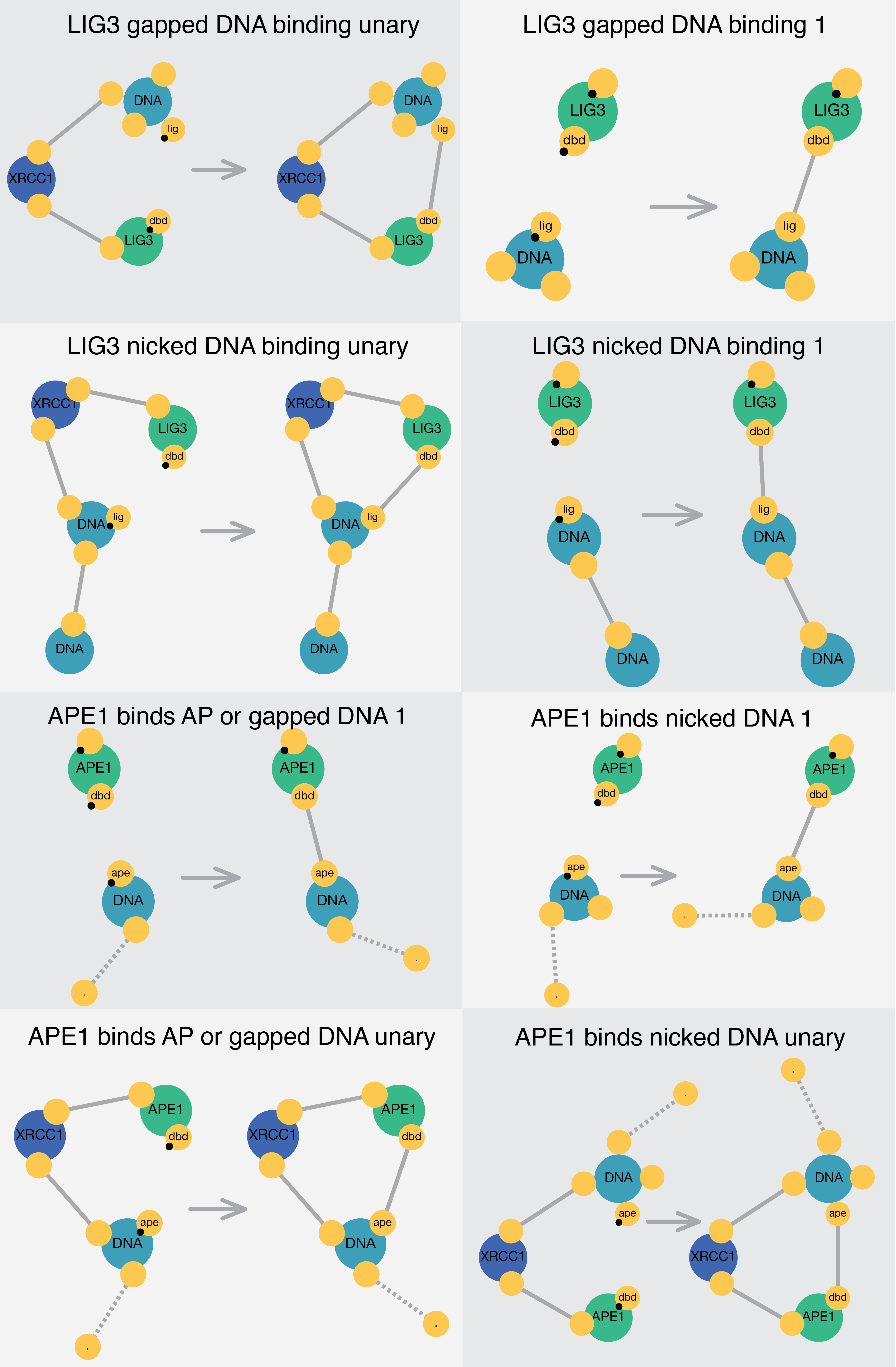}
      \caption{Subset of rules for DNA repair proteins DNA Ligase 3 (LIG3) and Apurinic/Apyrimidinic Endodeoxyribonuclease 1 (APE1), as described in the usage scenario in Sec.~\ref{s:usagescenario}. LIG3 is responsible for repairing nicks and gaps in DNA molecules. 
      Similarly to LIG3, APE1 is also responsible for repairing nicks and gaps along a double stranded DNA molecule, and is also able to repair apurinic sites in coordination with other proteins. Site names are only shown where changes occur to highlight differences between sides.}
    \label{fig:DNAa}
\end{figure}

\section{Usage Scenario}
\label{s:usagescenario}











Here, we provide brief examples in modeling DNA repair mechanisms, using \textit{RuleVis} to understand each step of creating salient rules for this model. DNA molecules in a living cell are subject to mutation through contact with other chemicals or environmental radiation. To regulate this process, organisms have several different types of repair proteins, which can vary species to species. Models of the functionality of different repair mechanisms are used to understand the behavior of various diseases, such as cancer, which can result from mutations to repair protein coding regions of the genome~\cite{cooper2017molecular}.

Fig.~\ref{fig:DNAa} shows sets of interactions for two specific DNA repair proteins, DNA ligase 3 (LIG3) and Apurinic/Apyrimidinic Endodeoxyribonuclease 1 (APE1). LIG3 is responsible for repairing nicks and gaps in double stranded DNA molecules. It is present in both the mitochondria and nuclei of human cells, catalyzing the reaction for reconnecting DNA during base excision repair, along with its cofactor protein XRCC1 ~\cite{kukshal2015human}. A ``nick'' is a discontinuity between two nucleotides in a double stranded DNA molecule, where there is no phosphodiester bond between them. Nicks can be created intentionally during cellular regulation, or as a stochastic process due to environmental conditions~\cite{cherepanov2002dynamic}. A ``gap'' is a missing sequence of nucleotides on one strand of double stranded DNA molecule~\cite{langston2006dna}. The protein APE1 repairs nicks and gaps in the DNA as well, but it also acts as the major apurinic/apyrimidinic (AP) endonuclease in human cells. In general, APE1 works in conjunction with other enzymes to repair damaged or missing purine bases on a DNA molecule~\cite{tell2009many}.

These example rules were originally developed over several years to create simulations of the DNA repair process. Our figures demonstrate that \textit{RuleVis} can quickly and reproducibly visualize existing executable rules, and use them to generate publication-ready figures. \textit{RuleVis} can accelerate the iterative scientific process of rule construction and evaluation, supporting the refining of rules for biological accuracy, promoting scholarly discussion and peer review, and fostering engagement with members of the broader biological research community who may not be familiar with rule-based modeling.

\section{Conclusion}


\textit{RuleVis} is an effective tool for constructing patterns and rules, and we continue to work with our systems biology collaborators to further integrate it into the Kappa software ecosystem. For future work, we plan to include additional style options, including a range of color, size, and glyph customization, as well as additional layout choices, making \textit{RuleVis} more useful for generating publication-ready figures. Furthermore, we will explore the use of visual highlighting to emphasize changes between the two sides of a rule, as well as to indicate mappings between the visualization and text representation. \textit{RuleVis} is available via our open source GitHub code repository at \url{https://github.com/CreativeCodingLab/RuleVis}, along with source code, instructions, and additional documentation. 








\bibliographystyle{abbrv-doi}

\bibliography{main.bib}

\begin{thebibliography}{10}

\bibitem{bostock2011d3}
M.~Bostock, V.~Ogievetsky, and J.~Heer.
\newblock D$^3$ data-driven documents.
\newblock {\em IEEE Transactions on Visualization and Computer Graphics},
  17(12):2301--2309, 2011.

\bibitem{kappamanual}
P.~Boutillier, J.~Feret, J.~Krivine, and W.~Fontana.
\newblock Kappa reference manual.
\newblock \url{https://kappalanguage.org/documentation}, accessed June 2019.

\bibitem{boutillier2018kappa}
P.~Boutillier, M.~Maasha, X.~Li, H.~F. Medina-Abarca, J.~Krivine, J.~Feret,
  I.~Cristescu, A.~G. Forbes, and W.~Fontana.
\newblock The {K}appa platform for rule-based modeling.
\newblock {\em Bioinformatics}, 34(13):i583--i592, 2018.

\bibitem{cherepanov2002dynamic}
A.~V. Cherepanov and S.~de~Vries.
\newblock Dynamic mechanism of nick recognition by {DNA} ligase.
\newblock {\em European Journal of Biochemistry}, 269(24):5993--5999, 2002.

\bibitem{chylek2015modeling}
L.~A. Chylek, L.~A. Harris, J.~R. Faeder, and W.~S. Hlavacek.
\newblock Modeling for (physical) biologists: {A}n introduction to the
  rule-based approach.
\newblock {\em Physical Biology}, 12(4):045007, 2015.

\bibitem{cooper2017molecular}
G.~M. Cooper and R.~Hausman.
\newblock {\em The cell: A molecular approach}.
\newblock Sinauer Associates, 2017.

\bibitem{biovisDang2}
T.~Dang, P.~Murray, J.~Aurisano, and A.~G. Forbes.
\newblock {ReactionFlow: A}n interactive visualization tool for causality
  analysis in biological pathways.
\newblock {\em {BMC Proceedings}}, 9(6):S6, 2015.

\bibitem{Dang2016a_IEEEEuroVis}
T.~Dang, N.~Pendar, and A.~G. Forbes.
\newblock {TimeArcs: V}isualizing fluctuations in dynamic networks.
\newblock {\em {Computer Graphics Forum}}, 35(3):61--69, June 2016.

\bibitem{danos2012graphs}
V.~Danos, J.~Feret, W.~Fontana, R.~Harmer, J.~Hayman, J.~Krivine,
  C.~Thompson-Walsh, and G.~Winskel.
\newblock Graphs, rewriting and pathway reconstruction for rule-based models.
\newblock In {\em LIPIcs-Leibniz International Proceedings in Informatics},
  vol.~18. Schloss Dagstuhl-Leibniz-Zentrum fuer Informatik, 2012.

\bibitem{danos09}
V.~Danos, J.~Feret, W.~Fontana, R.~Harmer, and J.~Krivine.
\newblock Rule-based modelling and model perturbation.
\newblock {\em Transactions on Computational Systems Biology}, XI:116--137,
  2009.

\bibitem{danos10}
V.~Danos, J.~Feret, W.~Fontana, R.~Harmer, and J.~Krivine.
\newblock Abstracting the differential semantics of rule-based models: Exact
  and automated model reduction.
\newblock In {\em Proceedings of the IEEE Symposium on Logic in Computer
  Science (LICS)}, pp. 362--381, 2010.

\bibitem{deeds12}
E.~J. Deeds, J.~Krivine, J.~Feret, V.~Danos, and W.~Fontana.
\newblock Combinatorial complexity and compositional drift in protein
  interaction networks.
\newblock {\em PLoS ONE}, 7(3):e32032, 2012.

\bibitem{Dwyer_COLA}
T.~{Dwyer}, Y.~{Koren}, and K.~{Marriott}.
\newblock {IPSep-CoLa: A}n incremental procedure for separation constraint
  layout of graphs.
\newblock {\em IEEE Transactions on Visualization and Computer Graphics},
  12(5):821--828, 2006.

\bibitem{earley1970efficient}
J.~Earley.
\newblock An efficient context-free parsing algorithm.
\newblock {\em Communications of the ACM}, 13(2):94--102, 1970.

\bibitem{fabregat2015reactome}
A.~Fabregat, K.~Sidiropoulos, P.~Garapati, M.~Gillespie, K.~Hausmann, R.~Haw,
  B.~Jassal, S.~Jupe, F.~Korninger, S.~McKay, et~al.
\newblock The reactome pathway knowledgebase.
\newblock {\em Nucleic Acids Research}, 44(D1):D481--D487, 2015.

\bibitem{forbes2017dynamic}
A.~G. Forbes, A.~Burks, K.~Lee, X.~Li, P.~Boutillier, J.~Krivine, and
  W.~Fontana.
\newblock Dynamic influence networks for rule-based models.
\newblock {\em IEEE Transactions on Visualization and Computer Graphics},
  24(1):184--194, 2017.

\bibitem{gehlenborg2010visualization}
N.~Gehlenborg, S.~I. O'donoghue, N.~S. Baliga, A.~Goesmann, M.~A. Hibbs,
  H.~Kitano, O.~Kohlbacher, H.~Neuweger, R.~Schneider, D.~Tenenbaum, et~al.
\newblock Visualization of omics data for systems biology.
\newblock {\em Nature Methods}, 7(3s):S56, 2010.

\bibitem{goel2009structure}
A.~K. Goel, S.~Rugaber, and S.~Vattam.
\newblock Structure, behavior, and function of complex systems: The structure,
  behavior, and function modeling language.
\newblock {\em AI EDAM}, 23(1):23--35, 2009.

\bibitem{harris2016bionetgen}
L.~A. Harris, J.~S. Hogg, J.-J. Tapia, J.~A. Sekar, S.~Gupta, I.~Korsunsky,
  A.~Arora, D.~Barua, R.~P. Sheehan, and J.~R. Faeder.
\newblock {BioNetGen 2.2: A}dvances in rule-based modeling.
\newblock {\em Bioinformatics}, 32(21):3366--3368, 2016.

\bibitem{kitano2002computational}
H.~Kitano.
\newblock Computational systems biology.
\newblock {\em Nature}, 420(6912):206, 2002.

\bibitem{kitano2005using}
H.~Kitano, A.~Funahashi, Y.~Matsuoka, and K.~Oda.
\newblock Using process diagrams for the graphical representation of biological
  networks.
\newblock {\em Nature Biotechnology}, 23(8):961, 2005.

\bibitem{kohl2011cytoscape}
M.~Kohl, S.~Wiese, and B.~Warscheid.
\newblock Cytoscape: {S}oftware for visualization and analysis of biological
  networks.
\newblock In {\em Data Mining in Proteomics}, pp. 291--303. Springer, 2011.

\bibitem{kolpakov2006biouml}
F.~Kolpakov, M.~Puzanov, and A.~Koshukov.
\newblock {BioUML: V}isual modeling, automated code generation and simulation
  of biological systems.
\newblock In {\em Proceedings of The Fifth International Conference on
  Bioinformatics of Genome Regulation and Structure}, pp. 281--284, 2006.

\bibitem{kukshal2015human}
V.~Kukshal, I.-K. Kim, G.~L. Hura, A.~E. Tomkinson, J.~A. Tainer, and
  T.~Ellenberger.
\newblock Human {DNA} ligase {III} bridges two {DNA} ends to promote specific
  intermolecular {DNA} end joining.
\newblock {\em Nucleic Acids Research}, 43(14):7021--7031, 2015.

\bibitem{langston2006dna}
L.~D. Langston and M.~O'Donnell.
\newblock {DNA replication: K}eep moving and don't mind the gap.
\newblock {\em Molecular Cell}, 23(2):155--160, 2006.

\bibitem{le2009systems}
N.~Le~Novere, M.~Hucka, H.~Mi, S.~Moodie, F.~Schreiber, A.~Sorokin, E.~Demir,
  K.~Wegner, M.~I. Aladjem, S.~M. Wimalaratne, et~al.
\newblock The systems biology graphical notation.
\newblock {\em Nature Biotechnology}, 27(8):735, 2009.

\bibitem{moody2008evaluating}
D.~Moody and J.~van Hillegersberg.
\newblock Evaluating the visual syntax of {UML: A}n analysis of the cognitive
  effectiveness of the {UML} family of diagrams.
\newblock In {\em Proceedings of the International Conference on Software
  Language Engineering}, pp. 16--34. Springer, 2008.

\bibitem{MurrayBMC2017}
P.~Murray, F.~McGee, and A.~G. Forbes.
\newblock A taxonomy of visualization tasks for the analysis of biological
  pathway data.
\newblock {\em {BMC Bioinformatics}}, 18(2):21--1--13, 2017.

\bibitem{sekar2017automated}
J.~A.~P. Sekar, J.-J. Tapia, and J.~R. Faeder.
\newblock Automated visualization of rule-based models.
\newblock {\em PLoS Computational Biology}, 13(11):e1005857, 2017.

\bibitem{shannon2003cytoscape}
P.~Shannon, A.~Markiel, O.~Ozier, N.~S. Baliga, J.~T. Wang, D.~Ramage, N.~Amin,
  B.~Schwikowski, and T.~Ideker.
\newblock {Cytoscape: A} software environment for integrated models of
  biomolecular interaction networks.
\newblock {\em Genome Research}, 13(11):2498--2504, 2003.

\bibitem{smith2012rulebender}
A.~M. Smith, W.~Xu, Y.~Sun, J.~R. Faeder, and G.~E. Marai.
\newblock {RuleBender: I}ntegrated modeling, simulation and visualization for
  rule-based intracellular biochemistry.
\newblock {\em BMC Bioinformatics}, 13(8):S3, 2012.

\bibitem{tell2009many}
G.~Tell, F.~Quadrifoglio, C.~Tiribelli, and M.~R. Kelley.
\newblock The many functions of {APE1/Ref-1: Not only a DNA} repair enzyme.
\newblock {\em Antioxidants \& Redox Signaling}, 11(3):601--619, 2009.

\bibitem{wilson2015kappa}
J.~Wilson-Kanamori, V.~Danos, T.~Thomson, and R.~Honorato-Zimmer.
\newblock Kappa rule-based modeling in synthetic biology.
\newblock {\em Computational Methods in Synthetic Biology}, pp. 105--135, 2015.

\end{thebibliography}
\end{document}